\documentclass[reprint, amsmath,amssymb, aps,pra,floatfix]{revtex4-2}

\usepackage{graphicx}
\usepackage{dcolumn}
\usepackage{bm}
\usepackage[position=t,singlelinecheck=off,justification=raggedright,caption=false]{subfig}
\usepackage[]{natbib}
\usepackage[colorlinks,citecolor=blue]{hyperref}
\usepackage{orcidlink}

\begin{document}

\title{Dirac-Rashba fermions and quantum valley Hall insulators\\ in graphene-based 2D heterostructures}
\author{Bo-Wen Yu}
\affiliation{Beijing National Laboratory for Condensed Matter Physics, Institute of Physics, Chinese Academy of Sciences, Beijing 100190, China}
\affiliation{School of Physical Sciences, University of Chinese Academy of Sciences, Beijing 100049, China}
\author{Bang-Gui Liu\orcidlink{0000-0002-6030-6680}}\email{bgliu@iphy.ac.cn}
\affiliation{Beijing National Laboratory for Condensed Matter Physics, Institute of Physics, Chinese Academy of Sciences, Beijing 100190, China}
\affiliation{School of Physical Sciences, University of Chinese Academy of Sciences, Beijing 100049, China}

\date{\today}

\begin{abstract}
It is highly desirable to modify and improve the Dirac electron system of graphene for novel electronic properties and promising applications. For this purpose, we study 2D heterostructures consisting of graphene and monolayer TMDs by means of first-principles calculation and effective low-energy hamiltonian model. We determine the model parameters by fitting with the first-prinples bands.   MoSe$_2$ and WSe$_2$ are chosen in order to align the Dirac cones of graphene with the intrinsic Fermi levels of the TMDs. It is found that the Dirac energy bands of graphene are modified, but the linear band dispersion near the cones is kept. It is shown that the effective low-energy model hosts Dirac-Rashba feimions in the WSe$_2$/graphene and MoSe$_2$/graphene/WSe$_2$, and there is quantum valley Hall effect in these graphene-based 2D heterostructures. Our further analyses indicate that there are strong interactions between the orbitals and spins especially near the K and K' points. These can be useful in further exploration  for novel properties and more functionalities in  2D heterostructures.
\end{abstract}

\maketitle

\section{Introduction}

Since the advent of graphene in 2004 \cite{rx00,rx01,rx02}, two-dimensional materials have been extensivly explored for their important phenomena and practical applications. Transition metal dichalcogenides (TMDs) as 2D semiconductors are well known for their unique qualities in scientific research and engineering applications \cite{rx03,rx04,rx05}. It is known that a combination of these two kinds of famous 2D materials can lead to much more rich and amazing properties. Actually, there have been many experimental achievements of heterostructures of graphene and monolayer TMDs thanks to the great development of experimental technology and methods \cite{rx06,rx07,rx08,rx09,rx10,rx11,rx12,rx13,rx14,rx15,rx16,rx17,rx18,rx19,rx20,rx21}. 

The graphene is the ideal 2D material for studying the Dirac feimions and topological properties in two dimensions becuase of its multiple degrees of freedom. Its orbital, spin, and valley features can be combined to form various interesting properties and topological structures\cite{qshe,qshe1,rx01,rpp2017,rx41,rx22}. However the tiny spin-orbit coupling (SOC) in pure graphene causes some limitations in various aspects. It is believed that some TMDs can be added to form graphene-based 2D heterostructures with strong SOC. As for important effect of SOC, Rashba effect near band edges of semiconductors is vital to determining carrier functionalities in semiconductor technology\cite{rashba,rashba1,rashba2003,rashba2,rashba3}. Considering recent experimetal advance \cite{rx01,rpp2017,rx41,rashba2,rashba3,science2024}, it is highly desirable to explore new phenomena and novel effects induced by the Dirac and Rashba features.

Here, we investigate 2D heterostructures consisting of graphene and monolayer TMDs (MoSe$_2$ and WSe$_2$) by means of first-principles methods and effective $k\cdot p$ low-energy models. These two TMDs are chosen in order to align the Dirac cones of graphene with their semiconductor gaps. We study the electronic structures of their optimized structures and construct a unified effectively low-energy model for further investigation. The parameetrs of the model are determined by fitting with the first-principles bands. It is found that the linear band dispersion is kept in a wide energy window and there are some modification in the bands, which leads to Dirac-Rashba feimions in  some of the heterostructures. We also calculate their Berry curvatures and find that the effective model hosts quantum valley Hall effect for some of the heterostructures. The detailed data and further analyses will be presented in the following.

\section{Methodology}

The first-principles calculations are performed with the projector-augmented wave (PAW) method within the density functional theory\cite{rx23}, as implemented in the Vienna Ab-initio simulation package software (VASP) \cite{rx24}. The generalized gradient approximation (GGA)  by Perdew,Burke, and Ernzerhof (PBEs)\cite{rx25} is used as the exchange-correlation functional. Our computational supercell includes $4\times 4$ primitive cells of monolayer graphene and $3\times 3$ primitive cells of H-TMD monolayers (the detailed structures and parameters will be presented in the following). The  self-consistent calculations are carried out with a $\Gamma$-centered ($4\times 4\times 1$) Monkhorst-Pack grid of the supercell\cite{rx26}. The kinetic energy cutoff of the plane wave is set to 450 eV. The convergence criteria of the total energy and force are set to 10$^{-6}$ eV and 0.01 eV/\AA{}. The spin-orbit coupling (SOC) is taken into account in the calculation of band structures and optimization of lattice structures. The inter-layer vacuum thickness is set to at least 30 \AA{} . The dipole correction \cite{rx27,rx28} is included in the calculation for the heterostructures. Dispersion corrections are taken into account via the Grimme approximation (DFT-D3)\cite{rx29}.

\section{Result and Discussion}

\begin{figure}[h]
\includegraphics[width=0.9\columnwidth]{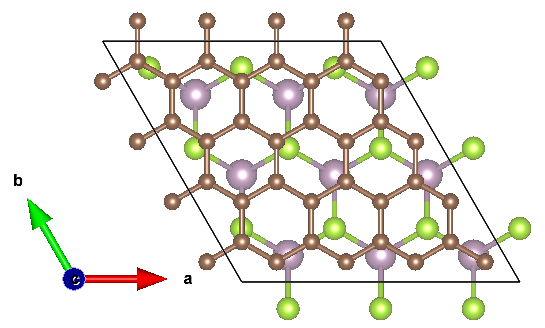}
\caption{\label{fig1} The structure of MoSe$_2$/graphene (WSe$_2$/graphene) heterostructure. Carbon, Mo (W), and Se atoms are brown , pink, and green, respectively.}
\end{figure}

The graphene and 2D TMDs structures have been extensively studied experimentally and theoretically for their interesting physical properties and applications\cite{rx30,rx21,rx20,rx31,rx32,rx33,rx30,rx34,rx35,rx36,rx37,rx38,rx39,rx40,rx41,rx42,rx43,rx44,rx45,science2024}. It was shown that the moir\'e pattern of a heterostructure can be eliminated by annealing process at a high temperature\cite{rx21}, which means that high stability can be achieved in some heterostructures without moir\'e pattern. We can construct some 2D heterostructures by combining graphene and appropriate H-phase TMD monolayers, in order to  achieve interesting 2D electronic structures and topological features. It is expected that while the Dirac cones of graphene will be changed by TMD monolayers, the main important features of graphene will be kept. We choose MoSe2 and WSe2 as TMD monolayers because the Dirac points (modified) are located approximately in the middle of their semiconductor gaps.  Consequently, we can realize MoSe$_2$/graphene, WSe$_2$/graphene, and MoSe$_2$/graphene/WSe$_2$ as our 2D heterostructures. In addition, we construct two symmetrical trilayer heterostructures for comparison: WSe$_2$/graphene/WSe$_2$ and MoSe$_2$/graphene/MoSe$_2$. The two bilayer heterostructures are described in Fig. \ref{fig1}, and the trilayer heterostructures are shown in Fig. s1 (supplemental materials). We also investigate another structure of WSe$_2$/graphene to show that the electronic structure is not sensitive to variation of inter-layer structure. This structure, WSe$_2$/graphene(1), can be constructed by removing the MoSe$_2$ monolayer in the trilayer MoSe$_2$/graphene/WSe$_2$.

The computational models contain one monolayer graphene and one or two monolayers of TMDs. It is necessary to reconcile the different lattice lengths of 4$\times$4 graphene cells and 3$\times$3 TMD cells. Graphene has bond length $a=1.42$ \AA{} and lattice constant $a_1=a_2=2.46 $ \AA{} \cite{rx01}. Both MoSe$_2$ and WSe$_2$ has $a^\prime_1=a^\prime_2=3.30 $ \AA{} \cite{rx30}, and then $3a^\prime_1/(4a_1)=1.006$. We apply a biaxial strain $0.6$\% to graphene to remove the small mismatch. Such a small strain can cause only a tiny change in the electronic structure of graphene \cite{rpp2017}. It is clear that there is no twisting in these 2D heterostructures.
In order to effectively perform structural optimization, it is helpful to choose a little smaller initial value for the inter-layer distance between graphene monolayer and TMD monolayer. The optimized distance value is  3.48 \AA{} for MoSe$_2$/graphene, 3.51 \AA{} for WSe$_2$/graphene, 3.46 \AA{} for MoSe$_2$/graphene/WSe$_2$, 3.42 \AA{} for WSe$_2$/graphene/WSe$_2$, or 3.44 \AA{} for MoSe$_2$/graphene/MoSe$_2$.


\begin{figure}[htbp]
\includegraphics[width=\columnwidth]{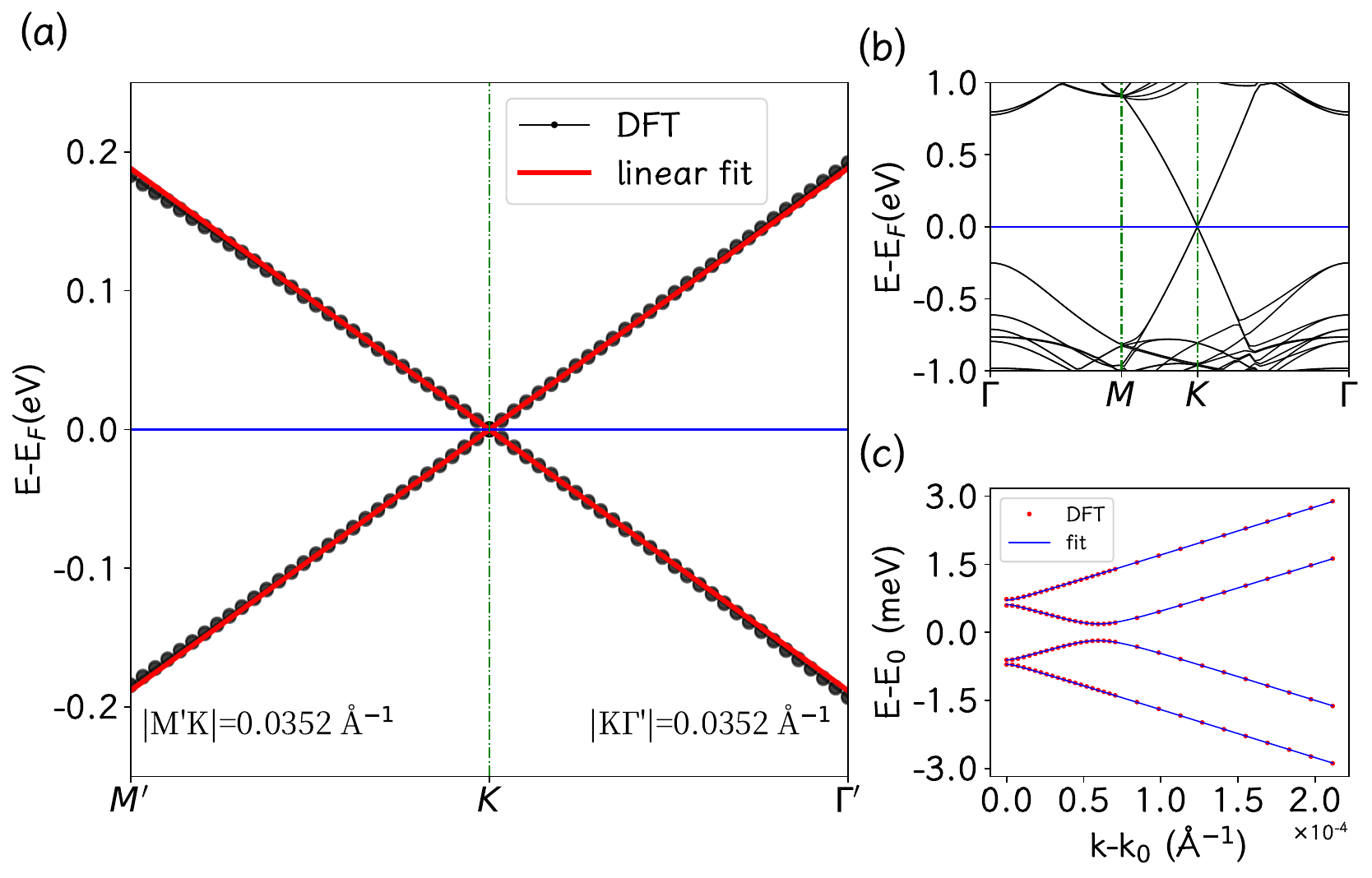}
\caption{\label{fig2} The electronic band structure of MoSe$_2$/graphene /WSe$_2$ heterostructure. (a) the energy bands near the K point to show the graphene-like linear dispersion between -0.25 and 0.25 eV, where the black lines describe the results of DFT calculation and the red lines the linear fitting. (b) The band structure of the 2D heterostructure in a wide energy window between -1 and 1 eV, showing the band features from the graphene and the 2D TMDs. (c) The band structure between -3 and 3 meV, demonstrating the Rashba effect near the Fermi level, where the red points are from DFT calculation and the blue lines the result of the fitting of the bands of the low energy model. The horizontal axis is the distance from the K point in the Brillouin zone.}
\end{figure}

With the optimized structures, we can investigate their electronic structures. The band structure of MoSe$_2$/graphene/WSe$_2$ heterostructure is presented in Fig. \ref{fig2}. It is very interesting that the Dirac cone of graphene is located within the semiconductor gaps of monolayer MoSe$_2$ and WSe$_2$ and there is an ideal linear dispersion in the graphene-dominant bands within the energy window [-0.2 eV, 0.2 eV]. This energy window corresponds to a circle of rarius of 0.035 \AA{}$^{-1}$ around K (K') point in the Brillouin zone. Actually, the bands near the Dirac cone are modified by the TMD monolayers, as shown in Fig. 2(c). The modification is limited to the small region near K (K') point. The conduction band minimum (CBM) and the valence band maximum (VBM) are both moved a little bit away from K (K') point, forming a circle around K (K') point. Usually, such a band edge indicates some Rashba effect\cite{rashba,rashba2003,rashba2,rashba3}. The energy window of the band modification is between -1.5 and 1.5 meV. The band structures of other heterostructures are shown in Fig. S2. Both WSe$_2$/graphene and WSe$_2$/graphene(1) have similar band structures. The MoSe$_2$/graphene and the two symmetric trilayer heterostructures have overall similar band structures, but their CBB and  VBT are both at K (K') point.


As shown in Fig. 2 and Fig. s2, the whole band structures are complex, but the low-energy parts of the band structures are simple, featuring the Dirac cones and linear dispersions. It is clear that the low-energy parts are similar to the low-energy bands of monolayer graphene. In each case, there are four bands, two conduction bands and two valence bands, for the low-energy part, and there is a symmetry between the conduction bands and the valence bands. Therefore, we use the follwoing $k\cdot p$ Hamiltonian (\ref{eq1}) to describe the low-energy bands near K (K') point in all the cases.
\begin{equation}
    \begin{split}
        \hat{H} = & v_F(\hat{\sigma}_x k_x+ \hat{\tau}_z \hat{\sigma}_y k_y)  +  \frac{R}{2}(\hat{\sigma}_y \hat{s}_x - \hat{\tau}_z \hat{\sigma}_x \hat{s}_y) \\
            & +\frac{A+B}{2}\hat{\sigma}_z + \frac{A-B}{2} \hat{\tau}_z\hat{s}_z,
    \end{split}
    \label{eq1}
\end{equation}
where $\vec{k}=(k_x, k_y)$ describes the k-vector in the 2D Brioullin zone, $\hat{\vec{\sigma}}=(\hat{\sigma}_x, \hat{\sigma}_y, \hat{\sigma}_z)$ are Pauli matrixes for the sublattice orbitals, ($\hat{s}_x, \hat{s}_y, \hat{s}_z$) are Pauli matrixes describing spin, and $\hat{\tau}_z$ is used to describe the valleys (with eigenvale 1 for K and -1 for K'). $\hat{\vec{\sigma}}$ can be considered to describe a pseudo-spin. 
The first term describes the low-energy bands of pure graphene, the second term is the  relativistic spin-orbit effect due to the TMD monolayers, and the last two terms are used to cause energy gaps due to the interaction between graphene and TMD monolayers.  It is clear that $\hat{\tau}_z$ is conserved because of $[\hat{\tau}_z,\hat{H}]=0$.

We can derive the following energy bands by solving the eigen equation of Hamiltonian (\ref{eq1}), $\hat{H}(\vec{k})\psi(\vec{k})=E(\vec{k})\psi(\vec{k})$.
    \begin{equation}
        \begin{split}
         &   E(\bm{k})=    \pm\sqrt{\frac{E_1(\bm{k}) \pm \sqrt{E_2(\bm{k})}}{2}} \\
          &  E_1(\bm{k})=  R^2 + A^2+B^2 + 2v_F^2\bm{k}^2                      \\
          &  E_2(\bm{k})=  (R^2-A^2+B^2)^2 +4v_F^2[R^2+(A-B)^2]\bm{k}^2
        \end{split}
        \label{eq2}
    \end{equation}
It is clear that both $E_1(\bm{k})$ and $E_2(\bm{k})$ are always positive and there are four bands (indexed with 0, 1, 2, and 3 from bottom to top).

The parameters $v_F$, $R$, $A$, and $B$ can be determined by fitting the bands of the hamiltonian (\ref{eq1}) with the DFT-calculated low-energy bands. The fitting values are summarized  in Table \ref{tab1}. As expected, $v_F$ is very large for all the six heterostructures, but the other parameters are very small, with their absolute values being 1.086 meV  at most. It is also interesting that the symmetric trilayers have $R=0$ and much smaller $A$ and $B$.

\begin{table}
\caption{The fitting parameters of the Hamiltonian for the six 2D graphene-based heterostructures.} \label{tab1}
 \begin{ruledtabular}
 \begin{tabular}{ccccc}
 System    & $v_F$ (meV$\cdot$\AA{})  & $R$ (meV) & $A$ (meV) & $B$ (meV)  \\
 \colrule
                MoSe$_2$/Gr            & 5403.00             & -0.349          & 0.167           & 0.521   \\ \hline
                WSe2/Gr                    & 5430.22             & -0.629          & 0.352           & -1.014  \\ \hline
                WSe2/Gr(1)               & 5409.96             & -0.664          & 0.380           & -1.086   \\ \hline
                MoSe$_2$/Gr/WSe$_2$  & 5303.28      & -0.329          & 0.708           & -0.514     \\ \hline
                WSe$_2$/Gr/WSe$_2$   & 5364.21       & 0.000           & 0.090           & 0.121         \\ \hline
                MoSe$_2$/Gr/MoSe$_2$ & 5386.36      & 0.000           & 0.071           & 0.061  
\end{tabular}
\end{ruledtabular}
\end{table}

\begin{figure}[htbp]
\includegraphics[width=\columnwidth]{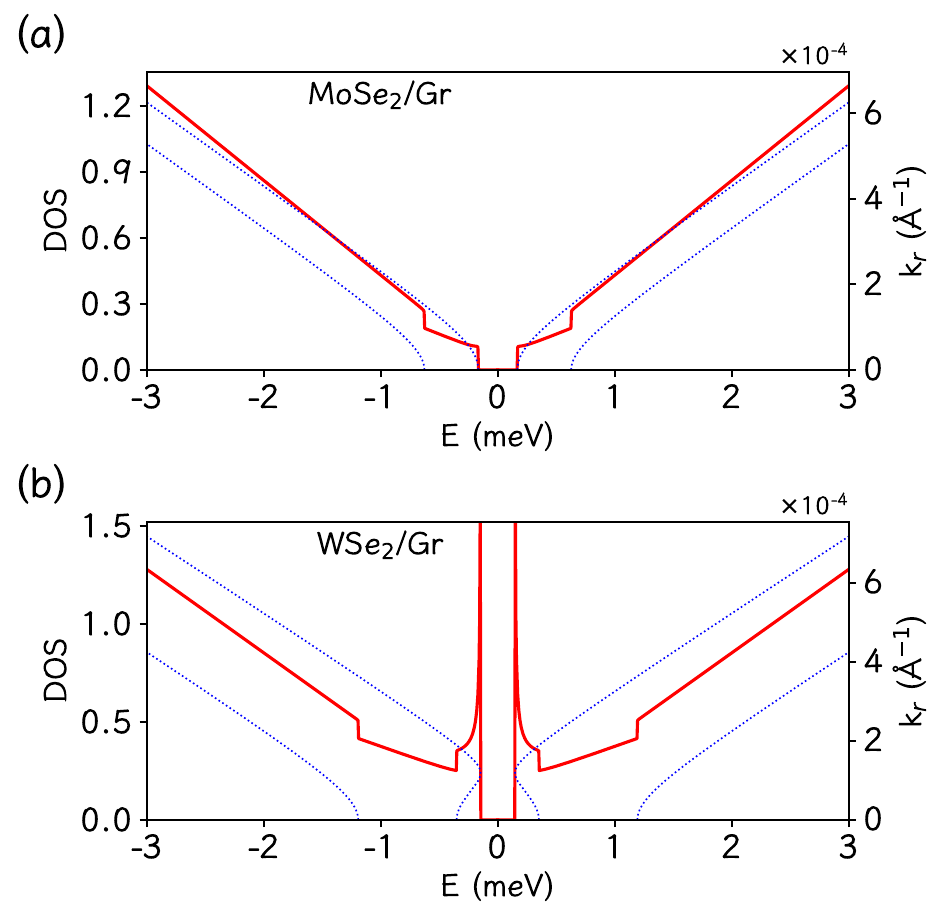}
\caption{\label{fig3} The density of states (DOS, red lines) of the MoSe$_2$/graphene and WSe$_2$/graphene  heterostructures for the corresponding energy bands (blue dots) in the energy window [-3meV, 3meV]. There exists a DOS divergence at both of the band edges in the case of WSe$_2$/graphene.}
\end{figure}

To give more information about the electronic structures, we present in Fig. \ref{fig3} density of states (DOS) with corresponding bands near CBM and VBM in the cases of MoSe$_2$/graphene and WSe$_2$/graphene. For MoSe$_2$/graphene, a DOS step is observed at each of the four band edges, which is the feature of two dimensions. In contrast, a big difference appears for WSe$_2$/graphene. A divergence of $1/\sqrt{E}$ is observed at both CBM and VBM, in addition to the two DOS steps from the other two bands. Actually, it can be shwon that there will be a circular band edge and then a band-edge divergence of $1/\sqrt{E}$ if the following condition is satisfied.
\begin{equation}
(A^2-AB)(B^2+R^2-AB)>0
\end{equation}
Such band edges implies that the Rashba effect plays some important role \cite{rashba,rashba2,rashba3,zsh-bitei}. Therefore, the effective low-energy model (\ref{eq1}) can host Dirac-Rashba feimions because the bands have linear dispersion in the wide windows in addition to the Rashba effect.  The parameters in Table \ref{tab1} implies that Dirac-Rashba feimions are formed in the WSe$_2$/graphene (and the variant) and MoSe$_2$/graphene.


In order to access topological properties, it is vecessary to calculate the Berry curvature of the energy bands. According to Kubo formula\cite{rx47}, the Berry curvature of the $n$-th band at the point $\vec{k}$, $\Omega^n(\vec{k})$, can be written as
\begin{equation}\label{eq:13}
\Omega^n=i\sum_{m\neq n}\frac{\langle n | \hat{v}_x|m\rangle \langle m |\hat{v}_y|n\rangle - \langle n | \hat{v}_y|m\rangle \langle m | \hat{v}_x|n\rangle}{(E_n-E_{m})^2}
\end{equation}
where $\hat{v}_x=\frac{\partial \hat{H}}{\partial k_x}$ and $\hat{v}_y=\frac{\partial \hat{H}}{\partial k_y}$, $E_n=E_n(\vec{k})$ is the $n$-th band anf $|n\rangle=|n\vec{k}\rangle$ (or $\psi_n(\vec{k})$) is the corresponding eigen wavefunction. The differential in the formula can be replaced with difference for the numerical calculation. Finally, the value of the Chern number can be obtained by summing the Berry curvature $\Omega^n(\vec{k})$ in terms of the symmetry of Hamiltonian (\ref{eq1}).

\begin{figure}[tbp]
\includegraphics[width=\columnwidth]{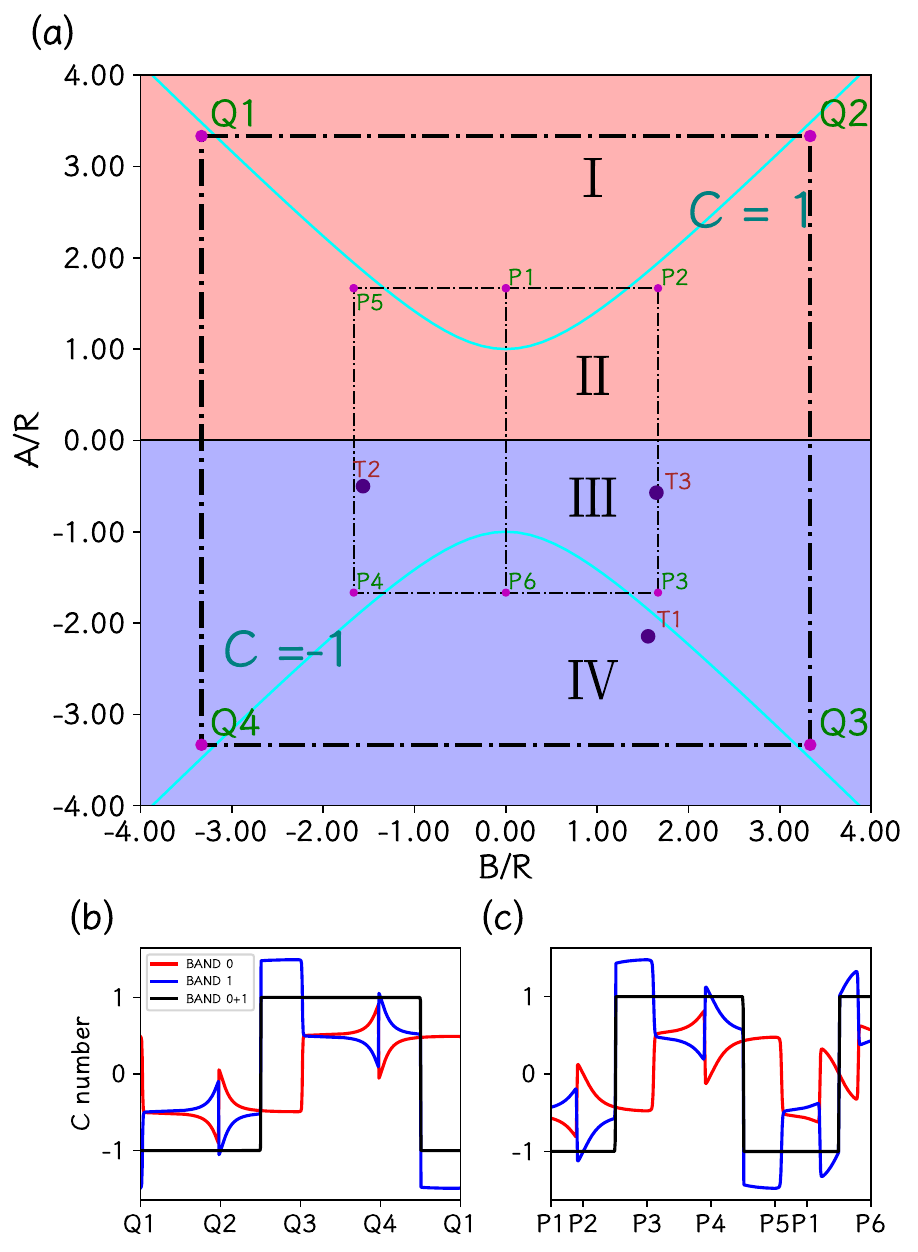}
\caption{\label{fig4} (a) The topological phase diagram for the K valley, with the Chern number $C=1$ ($C=-1$) in the upper (lower) half. The cyan line is defined by $(A/R)^2-(B/R)^2=1$. The point T1 is for the MoSe$_2$/graphene/WSe$_2$ heterostructure, T2 for the MoSe$_2$/graphene, T3 for the WSe$_2$/graphene.  (b) The band-resolved (red/blue)  and the total (black) Chern numbers  along path Q1-Q2-Q3-Q4-Q5-Q1 in the phase diagram. (c) The Chern numbers along P1-P2-P3-P4-P5-P1-P6. The red (blue) lines are for the lower (upper) valence band.}
\end{figure}

There are 4 effective parameters in this hamiltonian model, but three of them are important to the topological properties of this model because the parameter $v_F$ only determines the slope of the energy band near the K (K') point in the Brillouin zone. It is helpful to fix the parameter $R$  and scan the values of  $A$ and $B$ in the parameter space to calculate the topological Chern number, and then the topological phase diagram can be plotted with $A/R$ and $B/R$. The resulting phase diagram of Chern number ($C$) for the K valley and some band-resolved Chern numbers along two pathes  are presented in Fig. \ref{fig4}. Here $C$ is the sum of the two valence bands. It is clear in Fig. \ref{fig4}(a) that $C=1$ in the upper half-plane ($A/R>0$) and $C=-1$ in the lower half-plane ($A/R<0$), which implies that $A/R$ is the key parameter for the Chern number.  It is easy to show that  the semiconductor gap (at the Fermi level) is closed along the line $A=0$. The three points T1, T2, and T3 correspond to the three heterostructures:  MoSe$_2$/graphene/WSe$_2$, MoSe$_2$/graphene, and WSe$_2$/graphene. It should be pointed out that all the $C$ values are reversed for the K' valley because of time reversal symmetry. 

We can use $\hat{\tau}_z$ and the corresponding xy components to define a pseudo-spin operator for the valley degree of freedom, $\hat{\vec{\eta}}=(\hbar/2)\hat{\vec{\tau}}$, and thus $\hat{\eta}_z$ is also conserved. It is similar to the z component of the real spin operator.  $\eta_z=\hbar/2$ for the K valley, and  $\eta_z=-\hbar/2$ for the K' valley. Consequently, we obtain $C_\eta=(C_K-C_{K^\prime})/2=1$ for $A/R>0$ and  $C_\eta=(C_K-C_{K^\prime})/2=-1$ for $A/R<0$. Therefore, the model (\ref{eq1}) hosts a quantum valley Hall effect, similar to quantum spin Hall effect\cite{qshe,qshe1}. As shwon in Fig. 4, MoSe$_2$/graphene/WSe$_2$, MoSe$_2$/graphene, and WSe$_2$/graphene are in the region determined by $A/R<0$ and have $C_\eta=-1$. They can be considered to be quantum valley Hall insulators.

The phase diagram can be divided into four regions (\uppercase\expandafter{\romannumeral1}, \uppercase\expandafter{\romannumeral2}, \uppercase\expandafter{\romannumeral3}, \uppercase\expandafter{\romannumeral4}) by means of the curves $(A/R)^2-(B/R)^2=1$ and $A/R=0$. The two pathes are used to show the band-resolved Chern numbers, as presented in Fig. \ref{fig4} (b,c). The two valence bands (bands 0 and 1) have different contribution to the Chern number $C$, and one is larger than $C/2$ and the other smaller than $C/2$. It is interesting that the increments are opposite with respect to $C/2$ and will exchange sign when the curve $(A/R)^2-(B/R)^2=1$ is crossed. It can be shown from the band expression (\ref{eq2}) that the two valence bands at $\vec{k}=0$ (K, K') become degenerate when $(A/R)^2-(B/R)^2=1$ is satisfied.
When $|R|$ is small, $A/R$ and $B/R$ can be out of the parameter region in the phase diagram. Fortunately, we have shown that the phase diagram can be extended to large $A/R$ and $B/R$ (to $\pm\infty$).


\begin{figure}[tbp]
\includegraphics[width=0.49\textwidth]{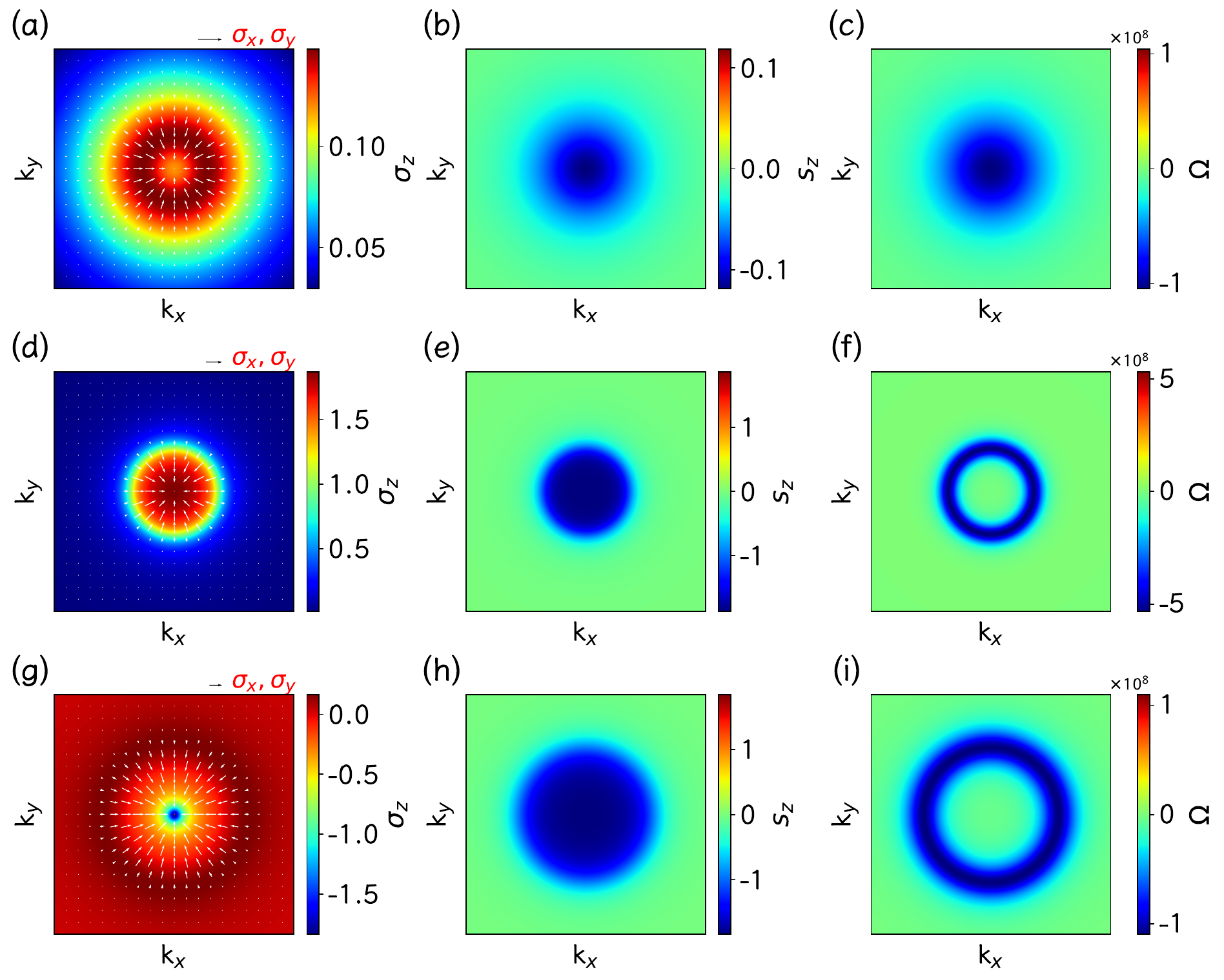}
\caption{\label{fig5} Schematic $k$-space distributions of orbital textures ($\Delta\vec{\sigma}$=($\sigma^0_x$-$\sigma^1_x$,$\sigma^0_y$-$\sigma^1_y$,$\sigma^0_z$-$\sigma^1_z$), left column), spin textures ($s_z$=$s^0_z$+$s^1_z$, middle column), and Berry curvatures ($\Omega$=$\Omega^0$+$\Omega^1$, right column) of the two valence bands (0 and 1) for MoSe$_2$/graphene (a,b,c), WSe$_2$/graphene (d,e,f) and MoSe$_2$/graphene/WSe$_2$ (g,h,i) as 2D heterostructures. The white arrows describe the direction and size of the xy components ($\Delta\sigma_x$,$\Delta\sigma_y$)  (a,d,g), and the color scales are used to indicate the $\Delta{\sigma}_z$ (a,d,g), $s_z$ (b,e,h), and $\Omega$ (c,f,i), respectively. }
\end{figure}

Furthermore, it is interesting to explore the interaction between the orbitals and the spins, and elucidate their relationships with the Berry curvature. We can calculate band-resolved average values of ($\hat{\sigma}_x, \hat{\sigma}_y, \hat{\sigma}_z$) and ($\hat{s}_x, \hat{s}_y, \hat{s}_z$) between the eigen-functions $|n\rangle=|n\vec{k}\rangle$. The average values of orbitals and spins can be written as
\begin{equation}
\vec{\sigma}^n(\vec{k})=\langle n\vec{k}|\hat{\vec{\sigma}}|n\vec{k}\rangle, ~~~
\vec{s}^n(\vec{k})=\langle n\vec{k}|\hat{\vec{s}}|n\vec{k}\rangle
\end{equation}
It is expected that $\vec{\sigma}^n(\vec{k})$ and $\vec{s}^n(\vec{k})$ will converge to those of monolayer graphene when the k-vector is far from the K (K') point, and the main defference will limited to the small regions near K (K') point (except some band translation). While the three band-resolved avarage values $\vec{\sigma}^n(\vec{k})$ are nonzero, $\Delta\vec{\sigma}(\vec{k})=(\sigma^0_x-\sigma^1_x,\sigma^0_y-\sigma^1_y,\sigma^0_z-\sigma^1_z)$ is nonzero only near the K (K') point. It is interesting that the xy components ($\Delta\sigma_x,\Delta\sigma_y$) remains in the radial direction in the xy plane. There are band-resolved helical spin textures $\vec{s}^n(\vec{k})$ due to the Rashba term, but the horizontal components ($s_x=s^0_x+s^1_x$ and $s_y=s^0_y+s^1_y$) of the two valence bands are equivalent to zero because the two valence bands have opposite contributions to the xy spin componennts, and then only the z component $s_z=s^0_z+s^1_z$ can be nonzero. To effectively show the relationship between the orbitals and the spins, we present in Fig. \ref{fig5} $\Delta\vec{\sigma}$ and $s_z$ of the MoSe$_2$/graphene, WSe$_2$/graphene, and MoSe$_2$/graphene/WSe$_2$. We focus on $\Delta\vec{\sigma}(\vec{k})$ and $s_z(\vec{k})$ because they are nonzero only near the K (K') point in the Brillouin zone. These correlated distributions near the K (K') point reflect the interaction between the orbials and the spins.

We also present Berry curvature $\Omega(\vec{k})=\Omega^0(\vec{k})+\Omega^1(\vec{k})$ near the K (K') point in Fig. \ref{fig5}. It is very interesting that the Berry curvature has main contribution near the band edges (the VBMs) and decays fast and diminishes to zero when the k-vector becomes far from the K (K') point. For  the MoSe$_2$/graphene, $\Omega(\vec{k})$ reaches its maximum at the K (K') point. In contrast, for the WSe$_2$/graphene  and the MoSe$_2$/graphene/WSe$_2$, $\Omega(\vec{k})$ takes its maximum along the circle around the K (K') point in the Brillouin zone  (defined by the valence band edge due to the Rashba effect).

\section{Conclusion}

In summary, six 2D heterostructures consisting of graphene and monolayer TMDs are investigated by means of first-principles calculation. We choose MoSe$_2$ and WSe$_2$ to align the Dirac cones of graphene with the intrinsic Fermi levels of the TMDs. It is found that the Dirac energy bands of graphene are modified by the TMDs, but the linear band dispersion near the K and K' points is kept in a wide energy window. We use an effective low-energy hamiltonian model to describe the energy bands and analyse the electronic properties, including possile topological feature. We determine the model parameters by fitting with the first-prinples bands. It is shown that the effective low-energy model hosts Dirac-Rashba feimions in the WSe$_2$/graphene and MoSe$_2$/graphene/WSe$_2$. Our model study also shows that there is quantum valley Hall effect in all the graphene-based 2D heterostructures. Our further analyses indicate that the orbitals and valleys can be described by the corresponding pseudo-spins, and there are strong interactions between the orbitals and spins especially near the K and K' points. These can be useful for exploring more properties and functionalities in 2D materials and 2D heterostructures for promising devices.

\begin{acknowledgments}
This work is supported by the Strategic Priority Research Program of the Chinese Academy of Sciences (Grant No. XDB33020100) and the Nature Science Foundation of China (Grant No.11974393).  All the numerical calculations were performed in the Milky Way \#2 Supercomputer system at the National Supercomputer Center of Guangzhou, Guangzhou, China.
\end{acknowledgments}

\bibliographystyle{apsrev4-2}

%

\end{document}